\RequirePackage{fix-cm}
\documentclass[smallextended]{svjour3}       
\smartqed  
\usepackage{graphicx}
\usepackage{amsmath}
\usepackage{epstopdf}
\usepackage{hyperref}
 \usepackage{mathptmx}      
\begin{document}
\title{Interfering Quantum Trajectories Without Which-Way Information}
\author{Kiran Mathew  \and
        Moncy V. John 
}


\institute{ K. Mathew \and M.V. John\at 
Department of Physics, St. Thomas College, Kozhencherri - 689641, Kerala, 					India \\
             M.V. John  \email{moncyjohn@yahoo.co.uk}\\
             K. Mathew  \email{kiran007x@yahoo.co.in} 
}

\date{Received: date / Accepted: date}

\maketitle

\begin{abstract}
Quantum trajectory-based descriptions of  interference between two coherent stationary waves in a double-slit experiment are presented, as given by  the de Broglie-Bohm (dBB) and modified de Broglie-Bohm (MdBB) formulations of quantum mechanics. In the dBB trajectory representation, interference between two spreading wave packets can  be shown also as resulting from motion of particles. But   a trajectory explanation for interference between stationary states is so far not available in this scheme. We show that both the dBB and MdBB trajectories are  capable of producing the interference pattern  for stationary  as well as wave packet states. However,  the dBB representation is found to  provide the  `which-way' information that helps to identify the hole through which the particle emanates. On the other hand, the MdBB representation does not provide any which-way information while  giving a satisfactory explanation of interference phenomenon in tune with the de Broglie's wave particle duality. By counting the trajectories reaching  the screen, we have numerically evaluated the intensity distribution of the fringes and found very good agreement with the standard results.
\end{abstract}

\maketitle

\section{Introduction}

The complementarity principle of Bohr \cite{bohr} states that a quantum system can  behave either as a system of particles or as a wave, but never simultaneously  as both. On the other hand, de Broglie's  wave-particle duality is less restrictive, since a `both-particle-and-wave' picture can also be accommodated in it.  In 1909, immediately after his explanation of photoelectric effect, Einstein attempted to describe  the motion of  the quanta of radiation (photons) as localised singular points  \cite{einstein3,einstein1}. Apparently, he wanted to keep both the wave and particle pictures and to present the  quantum theory of light as  Newton's corpuscular theory in some new form, thereby continuing with the age-old discussion on the nature of light.  However,  this  `photon as particle' concept did not enjoy much appreciation till recent times, for various reasons \cite{holland_book}.  In his epoch-making  work in 1923, de Broglie suggested that if radiation has both  wave and particle nature, matter  also has them.  Applying this principle of wave-particle duality \cite{db1,db2}, he not only predicted wave-like behaviour for beams of electrons etc. - his attempt was also to develop a new mechanics  \cite{valentini} by treating them as matter particles themselves.  Thus we see that in  de Broglie's  wave-particle duality perspective,  every physical system has wave and particle nature, so that one  can   describe it    using both   wave mechanics  and   particle mechanics.  In the latter case,  the mechanics obeyed by them may be  non-Newtonian and it was for this purpose that he developed the `pilot wave theory'  in the 1920's. But it  failed to get acceptance and was abandoned even by de Broglie himself for a long time. In 1952, the theory was revived by Bohm  \cite{bohm_1}. Afterwards,  Bohm and his collaborators, along with many others, succeeded in   `demystifying' several quantum phenomena using this trajectory approach. The formalism now provides one of the most attractive alternative interpretations of quantum mechanics and is   called the de Broglie-Bohm (dBB) quantum mechanics. Even Einstein's `photon as particle' concept has  received some renewed attention in recent times. The measurement of `weak-valued trajectories' \cite{kocsis} of single photons, as they undergo two-slit interference,  is reported to be identical to those predicted in the dBB interpretation of quantum mechanics. 

It may  be noted that dBB is not the only quantum trajectory formalism available in the literature. The Floyd, Faraggi and Matone (FFM) \cite{Floyd_1}  and the modified de Broglie-Bohn (MdBB) \cite{MdBB_MVJ_1,mvj_prob1,mvj_prob2,mvj_cohe,mvj_tunnel} trajectory representations   have also received wide attention in recent years.  The equations of motion used in dBB and MdBB schemes are alike, of the general form $m{\bf \dot{r_i}} =\nabla_i S$, where $S$ represents the Hamilton-Jacobi functions in the respective quantum Hamilton-Jacobi equations in the two schemes. But in the FFM  representation, a different equation of motion, rendered by Jacobi's theorem, is used. Another difference is that the dBB and FFM are trajectory representations in real space, but in MdBB, the trajectories lie in a complex space. In MdBB, the connection with the real world is established by postulating that the real part of the trajectories correspond to the physical trajectories \cite{MdBB_MVJ_1}. With regard to the use of probability, the three schemes differ in the following way. The dBB and the MdBB approaches use the same Born's probability axiom to make all statistical predictions and hence claim equivalence with standard quantum mechanics in all experimental situations. For instance, Holland \cite{holland_book} lists the Born probability axiom as a basic postulate of the dBB theory. In MdBB, the same axiom is followed to evaluate the mean values etc., where the integration is performed along the real line only \cite{MdBB_MVJ_1}. It is also worth mentioning that D$\ddot{u}$rr, Goldstein and Zanghi \cite{durr} have provided a justification for Born rule in dBB, such that it only describes the statistical regularities of systems in quantum equilibrium. On the other hand, MdBB is capable of providing an expression for $|\Psi|^2$ in terms of the velocity field along the real line \cite{mvj_prob1,mvj_prob2}.  The FFM representation claims not to involve probability at all.

It was found that  in bound state problems with  real wave functions, the dBB velocity of the particles turns out to be zero everywhere \cite{R_carroll_1}.  This  behaviour is counter-intuitive in a quantum theory of motion.
But the MdBB quantum mechanics, which   puts forward a new  dynamics based on a complex action, is found successful in this case \cite{MdBB_MVJ_1}. Similarly, in  the  tunneling of potential barriers,     when  the incident particle is described by a  stationary energy eigenfunction,  the  dBB trajectories are always proceeding towards the potential barrier and there are no reflected trajectories \cite{mvj_tunnel}.  The usual practice adopted in the dBB scheme to circumvent these is  to take a wave packet as the initial  wave function.  Then one draws the  trajectories for an ensemble of particles by integrating the equation of motion proposed by de Broglie. The starting points of these trajectories are chosen according to the $\Psi^{\star}\Psi$ distribution of the initial wave packet.  The  family of trajectories thus obtained help us to get the final distribution of particles and thereby to deduce the evolution of the wave packet. But it may be noted that  the continuity equation for $\Psi^{\star}\Psi$ in real space is `in-built' in the dBB scheme   and hence for the wave-packets, it is only natural that the non-crossing trajectories  evolve to a final distribution that coincides with the one predicted in standard quantum mechanics. On the other hand, in \cite{mvj_tunnel,chou_tnl_1,chou_tnl_2}, it was shown that the crossing  trajectories in the MdBB scheme (which also admits a continuity equation for $|\Psi|^2$ along the real space) can exhibit quantum tunneling through barriers, even for stationary states. 

 In this paper, we  check whether dBB and MdBB trajectory representations can produce  the same  interference  pattern as that  in  standard quantum wave mechanics, for certain suitable wave functions in the  double-slit experiment. Of our particular interest is the  case of stationary state wave functions. Recently it was reported that some experiments with  double slits    endorse the existence of dBB-like trajectories    \cite{kolasinski,jonsson,gondran,mahler}  showing quantum interference. But we may note that in these cases  too,   all attempts to draw  dBB trajectories  have  considered only wave packets for describing the  incident particles. We here  show that also for a stationary energy eigenstate, such as a dispherical wave function  emerging from two holes on a barrier, both the dBB and MdBB schemes can give the desired   pattern.   This success  is an important result,  for it takes the  schemes closer to the de Broglie's original idea of wave-particle duality.   In particular,  it highlights  that in both cases, as the particles move on,  trajectories condense to those regions where the probability density is high, in spite of the fact that  there is no flow of probability in the stationary state, as per the standard quantum mechanics.

However, in the dBB scheme, while using both the wave packet description   and  the stationary dispherical wave function,  we can identify the slit through which the particle has emerged, by knowing the position of particles on the screen (when it is detected). This `which-way' information is  an inescapable conclusion in the dBB approach and is considered as a success. But this is not compatible with the world-view of standard quantum mechanics  \cite{feynman}. We have drawn the trajectories also in the MdBB representation and found that the MdBB approach gives the desired  pattern without any which-way information.


\section{ Interference  in dBB quantum mechanics} \label{sec:dbb_wavep}

The original attempt to explain  interference   by drawing  particle trajectories  in a  double slit experiment was made by Philippidis et. al. \cite{holland_book,DBB_Phili}. For this, these authors considered the superposition of two waves, both of which  propagate as plane waves  towards the screen,  but at the same time  are spreading wave packets along a direction perpendicular to the slits. We make slight modifications to their experimental set-up and in all our examples consider interference of waves emanating from two holes made on a plane barrier.  The holes act as secondary sources.  Let us first reproduce the result in \cite{DBB_Phili}  to set our background and notations. Consider the two holes    on the barrier placed in the $yz$-plane at $x=0$. On the barrier, let the  centres of the holes  be at $z=\pm Z_0=\pm 10$.  In the wave packet case, the two holes   are assumed to be `soft', such that they  generate waves having identical Gaussian profiles along the $y$ and $z$-directions  at $t=0$. The interference pattern is obtained on a screen  placed parallel to the barrier, at $x=D=50$.  The two waves, emerging from the holes $A$ and $B$, can therefore be described by \cite{holland_book}

\begin{equation}
\begin{split}
\psi_A(x,y,z,t)=&\dfrac{C_A}  {(2\pi \sigma_t^2)^{{1}/{4}}}
\exp \left[\dfrac{-\left(z-Z_0\right)^2-y^2}{4\sigma_0 \sigma_t}\right]
\\
 &\times \exp \left[  i\left( k_x x-\dfrac{E_x t}{\hbar}\right)\right],
\end{split}
\end{equation} 
and

\begin{equation}
\begin{split}
\psi_B(x,y,z,t)=&\dfrac{C_B}  {(2\pi \sigma_t^2)^{{1}/{4}}}
 \exp \left[\dfrac{-\left(z+Z_0\right)^2-y^2}{4\sigma_0 \sigma_t}\right]
\\
 &\times \exp \left[  i\left( k_x x-\dfrac{E_x t}{\hbar}\right)\right],
\end{split}
\end{equation} 
respectively.   Here $E_x= \hbar^2k_x^2/(2m)$. Note that  $\psi_A$ is both a  propagating  plane wave  in the $x$-direction and  spreading wave packet along  the $y$ and $z$-directions. Similar is the case for the packet $\psi_B$. The packets  spread into one another. With $\sigma_0$ as the  initial value, the width of a wave packet at  time t is $\sigma_t=\sigma_0\left(1+i\dfrac{\hbar t}{2 m \sigma_0^2}\right)$. However,  the plane wave along the $x$-direction is unaffected. The total wave function in the region between the barrier and the screen is given by the superposition 

\begin{equation}
\psi(x,y,z,t)=\psi_A(x,y,z,t)+\psi_B(x,y,z,t). \label{eq:Wav_Fun}
\end{equation}
With $C_A=C_B$, this wave function is factorisable, for one can write it as $\psi(x,y,z,t)=f_1(x,t)f_2(y,t)f_3(z,t)$. Let $S$ be its phase  such that  in the polar form, we have $\psi=Re^{iS/\hbar}$. The trajectories in the dBB scheme are obtained by integrating the equation of motion \cite{holland_book}

\begin{eqnarray}
 \dfrac{d\mathbf{r}}{dt}=\dfrac{1}{m}\nabla S=-\dfrac{i \hbar}{2m} \left( \dfrac{\psi^* \nabla \psi- \psi\nabla \psi^* }{\psi^*\psi} \right)=Re\left(-\dfrac{i \hbar}{m}\dfrac{1}{\psi}\nabla \psi\right).
\label{eq:vel_dBB} 
\end{eqnarray} 

 Substituting the wave function (\ref{eq:Wav_Fun}) in   (\ref{eq:vel_dBB}), with $C_A=C_B$, we get \cite{holland_book}
 
\begin{equation} 
\dfrac{dx}{dt}=\dfrac{\hbar k_x}{m},
\label{Eq:Vel_dBB_fun_x} 
\end{equation}

\begin{equation} 
\dfrac{dy}{dt}=Re\left[ \frac{i\hbar}{m}\left(\frac{y}{2\sigma_0\sigma_t}\right)\right],
\label{Eq:Vel_dBB_fun_y} 
\end{equation}
and

\begin{equation}
\dfrac{dz}{dt}= Re\left\{\dfrac{i \hbar}{m}\left[ \dfrac{ z-Z_0 \tanh\left(\frac{z Z_0}{2 \sigma_0 \sigma_t}\right)}{2\sigma_0\sigma_t}\right]\right\} .
\label{Eq:Vel_dBB_fun_z} 
\end{equation}
 Here, $\hbar,\ k_x \ \text{and} \ m$ are real constants.  Integrating equation (\ref{Eq:Vel_dBB_fun_x}) gives
\begin{equation}
x(t)=\dfrac{\hbar k_x}{m} t+x_0.
\label{Eq:Vel_dBB_x}
\end{equation}
One notes that with $x_0=0$,  a plot of $y$ or $z$ versus $x$ will look the same as a  plot of $y$ or $z$, respectively, versus $t$. In the wave packet case, let us take $\hbar /m=1 $,   $k_x=1$, $\sigma_0=1$,  and $0\leq t \leq T$, with $T=50$ in all calculations. Energy $E_x$ is equal to $\hbar^2 k_x^2/(2m)=m/2$, with the above parameter values. When we choose $y_0=0$ as the  $y$-coordinate of the initial point of a trajectory, according to equation (\ref{Eq:Vel_dBB_fun_y}), that trajectory will remain confined to the $xz$-plane. Hence we can restrict ourselves to drawing  trajectories  in the two-dimensional $xz$-plane  by resorting to this condition. In this wave packet case, all trajectories are drawn from  starting points with $x$-coordinate as    $x_0=0$. Along the $z$-direction, we choose  equidistant  points in the interval $Z_0 -\delta \leq z \leq Z_0 +\delta$ and $-Z_0 -\delta \leq z \leq -Z_0 +\delta$. Equation (\ref{Eq:Vel_dBB_fun_z}) was solved numerically using fourth order Runge-Kutta  method. The step-size for the parameter $t$ used in our calculations was $\Delta t = 0.01$.

\begin{figure}[h]
\includegraphics[width=0.8\textwidth]{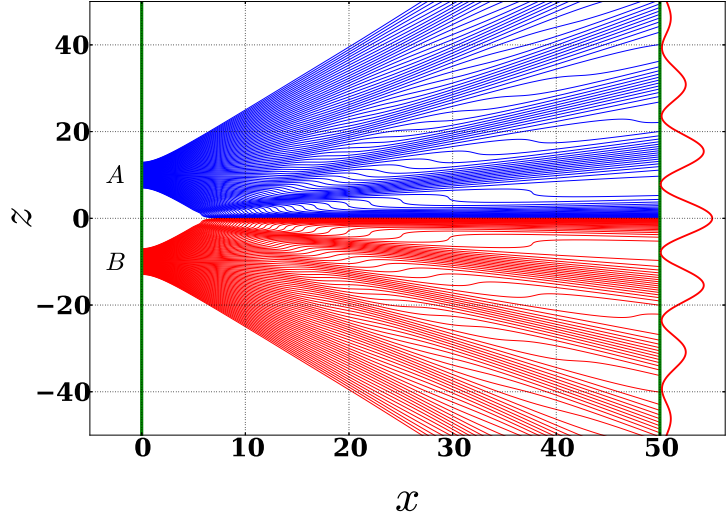}
\caption{dBB trajectories for two Gaussian wave packets with starting points at $x_0=0$, $y_0=0$ and $z$-coordinate uniformly distributed in the interval $Z_0-\delta <z_0<Z_0+\delta$ for hole $A$ and $-Z_0-\delta <z_0<-Z_0+\delta$ for hole $B$. The values $Z_0=10$ and $\delta =3$. Adjacent panel on the right side shows standard $\psi^{\star}\psi$ probability density in this case, in the direction of the $z$-axis, at $x=50$ and $y=0$ on the screen.}
\label{fig:dBB_wavepacket}        
\end{figure}

 The plot of dBB trajectories, with $Z_0=10$  and $\delta = 3$  is shown in Fig. \ref{fig:dBB_wavepacket}. They  are exactly of the form reported in \cite{DBB_Phili}. The adjacent panel shows the standard probability density along the $z$-axis, evaluated as per the $\psi^{\star}\psi$-distribution, at $x=50$ and $y=0$. It is now easy to see that the trajectory pattern has the same band width as that in standard quantum mechanics.

Though this pattern agrees with the interference bands predicted in the standard wave representation,   it exhibits the  feature  that the respective trajectories emanating from the  slits $A$ and $B$   never meet each other. Those  trajectories whose starting points are near the upper slit ($z>0$) cannot go to the region below some point  with $z=0$ on the screen and vice-versa. Thus there is a kind of fictitious barrier between the two regions, so that the two families of trajectories   appear to repel each other \cite{Floyd_1,dBB_ASZ_1,dBB_ASZ_2,dBB_ASZ_3,dBB_ASZ_4,dBB_ghose}.    Conversely, by knowing the point at which the particle reaches the screen, one can identify the slit through which it has emanated. This which-way information cannot be obtained in the standard interpretation of quantum mechanics.

 In general, at any given point and at a given time, the velocity of a particle on the dBB trajectory is definite and single-valued. Such simultaneous, well-defined  values for position and velocity, as can be evaluated using  equation (\ref{eq:vel_dBB}), are itself against the world view of standard quantum mechanics. One can see that  this single-valuedness  leads to the non-crossing property \cite{Floyd_1,dBB_ASZ_1,dBB_ASZ_2,dBB_ASZ_3,dBB_ASZ_4} of dBB trajectories. The consequent which-way information is an inescapable conclusion in the dBB representation.

\section{dBB trajectories for a stationary state} \label{sec:dbb_stat}
 
Instead of  wave packets, let us now consider the superposition of two stationary  spherical wave functions \cite{floyd}. These waves emanate from   two holes made on a plane barrier placed at $x=0$. The holes  are  a pair of  secondary point sources activated coherently by a primary source.  These secondary point sources emit  spherical waves that are components of the total dispherical wave function. As in the previous case, we treat the problem in Cartesian coordinates $(x,y,z)$ and the two holes are located at $(x =0, y=0, z=\pm Z_0 =\pm 10)$. With

\begin{equation}
r_1=\left[ x^2+y^2+(z-Z_0)^2\right]^{1/2}, \label{eq:rone}
\end{equation}
and
 
 \begin{equation}
r_2=\left[ x^2+y^2+(z+Z_0)^2\right]^{1/2},\label{eq:rtwo}
\end{equation}
the dispherical wave function can be written in the form  \cite{floyd}

\begin{equation}
\psi_d=\dfrac{\exp(i{\bf k_1}.{\bf r_1})}{r_1}+\dfrac{\exp(i{\bf k_2}.{\bf r_2})}{r_2} = \dfrac{\exp(i{ k}{ r_1})}{r_1}+\dfrac{\exp(i{ k}{ r_2})}{r_2}, 
\label{eq:dispherical}
\end{equation}
where ${\bf k_1}.{\bf k_1}={\bf k_2}.{\bf k_2}=k^2$.  For the dispherical wave function, we shall choose $\hbar/m=1$, $k=1$ in all calculations. The energy of these particles are again $E=\hbar^2 k^2/(2m)=m/2$. 

  Substituting for $r_1$ and $r_2$ from equations (\ref{eq:rone}) and (\ref{eq:rtwo}) respectively in equation (\ref{eq:dispherical}), and using equation (\ref{eq:vel_dBB}), one can write the dBB equations of motion in Cartesian coordinates. The partial derivatives to be used in these equations are, respectively,

\begin{equation}
\dfrac{\partial \psi_d}{\partial x}=\dfrac{x (i+k r_1) \exp{\left(i k r_1\right)}}{r_1^3}+\dfrac{ x  (i+kr_2)\exp{\left(i k r_2 \right)}}{r_2^3}, \label{eq:parpsid1}
\end{equation}

\begin{equation}
\dfrac{\partial \psi_d}{\partial y}=\dfrac{y (i+k r_1) \exp{\left(i k r_1\right)}}{r_1^3}+\dfrac{ y  (i+kr_2)\exp{\left(i k r_2 \right)}}{r_2^3}, \label{eq:parpsid3}
\end{equation}

and

\begin{equation}
\dfrac{\partial \psi_d}{\partial z}=\dfrac{(z-Z_0) (i+k r_1) \exp{\left(i k r_1\right)}}{r_1^3}+\dfrac{(z+Z_0)  (i+kr_2)\exp{\left(i k r_2 \right)}}{r_2^3}, \label{eq:parpsid2}
\end{equation}
where one substitutes for $r_1$ and $r_2$ from equations (\ref{eq:rone}) and (\ref{eq:rtwo}). 

\begin{figure}[h]
\includegraphics[width=0.8\textwidth]{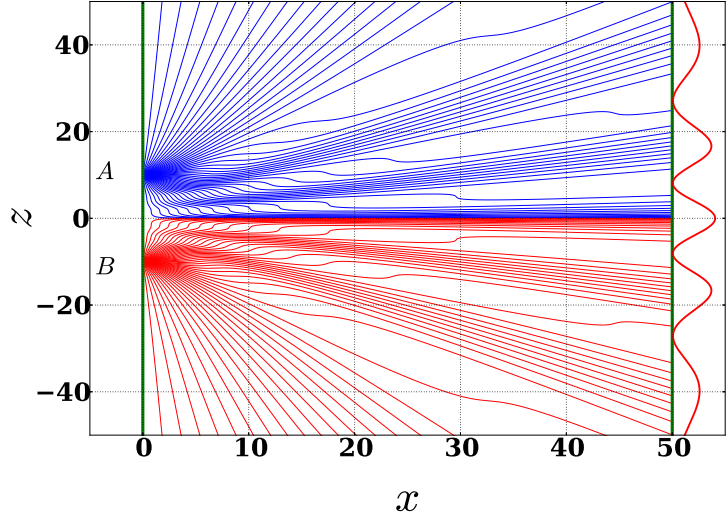}
\caption{dBB trajectories for stationary dispherical wave function, starting from initial points obeying $y_0=0$ and $x_0^2+z_0^2=a^2$, with $x_0>0$ and $a=10^{-3}$. They start from equidistant points lying on a semicircle of radius $a=10^{-3}$ surrounding the holes, and in the $xz$-plane. Adjacent panel on the right side shows standard $\psi^{\star}\psi$ probability density in this case, along the $z$-axis ($x=50$ and $y=0$) on the screen.}
\label{fig:dBB_stationary1}        
\end{figure}

As in the previous case, here also one can put $y_0=0$ as an initial condition for the variable $y$ and then effectively treat the problem in the two dimensional $xz$-plane. That this is possible can be verified from equation (\ref{eq:parpsid3}).
Also we note that the trajectories cannot start from the exact location of the holes,  with either $r_1=0$ or  $r_2=0$, because the wave function itself is infinite  at these points. To circumvent this difficulty, we shall begin the trajectories from equidistant points  on a semi-circle  of radius $a$ surrounding the holes and lying in the $xz$-plane and choose the radius $a$ as small as possible. The equations of motion are solved using Runge-Kutta fourth order method for $0<t<50$ and the step-size was $\Delta t=0.01$. The resulting trajectories for $a = 10^{-3}$ are shown in Fig. \ref{fig:dBB_stationary1}.

The adjacent panel shows the probability distribution in standard $\psi^{\star}\psi$ approach. There is  very  good agreement between the band width obtained from the trajectories and that in the standard approach.

One can see   that the  non-crossing property of the dBB trajectories   leads to   which-way information in the present case of the stationary state also. In the next sections, we shall see that the MdBB trajectories can exhibit the desired trajectory pattern, even when they can cross each other and have no which-way information. 

\section{Interference pattern in the MdBB approach - Wave packets}

When we write the quantum wave  function in the form $\psi= \exp\left({iS}/{\hbar}\right)$,  the Schrodinger equation becomes

\begin{equation}
\dfrac{1}{2m}\left(\nabla S\right)^2+V+\dfrac{\partial S}{\partial t}-\dfrac{i\hbar}{2m} \left(\nabla^2 S\right)=0 . \label{eq:qhje}
\end{equation}
This  is known as quantum Hamilton-Jacobi equation (QHJE).
 The complex quantum trajectories are obtained by integrating the MdBB equation of motion  \cite{MdBB_MVJ_1}
 
 \begin{equation}
\dfrac{d\mathbf{r}}{dt}=\nabla S=-\dfrac{i\hbar}{m} \dfrac{1}{\psi}\nabla \psi.
 \label{Eq:Vel_MdBB} 
 \end{equation}
 Note that the solution of this equation    leads to trajectories in a complex space with  coordinates $x=x_r+i x_i$, $y=y_r+iy_i$ and $z=z_r+i z_i$.
On the other hand, the  velocity field  given by the dBB equation of motion (\ref{eq:vel_dBB}) is defined only over the real space and is just  the real part of the above velocity $d{\bf r}/dt$. 

 First, let us draw the complex trajectories corresponding to a wave packet. Assuming the same experimental set up as described in Sec.  \ref{sec:dbb_wavep}, we use the wave function (\ref{eq:Wav_Fun})  in  equation (\ref{Eq:Vel_MdBB}) to get
 
\begin{equation}
\dfrac{dx}{dt} =\dfrac{\hbar k_x}{m},
\label{Eq:Vel_MdBB_fun_x} 
 \end{equation}

 \begin{equation}
\dfrac{dy}{dt} =\frac{i\hbar}{m}\; \frac{y}{2\sigma_0\sigma_t},
\label{Eq:Vel_MdBB_fun_y} 
 \end{equation}
and 

\begin{equation}
\dfrac{dz}{dt}=\dfrac{i \hbar}{m}\left[ \dfrac{z-Z_0 \tanh\left(\frac{z Z_0}{2 \sigma_0 \sigma_t}\right)}{2\sigma_0\sigma_t}\right], 
\label{Eq:Vel_MdBB_fun_z} 
\end{equation}  
where $\hbar$, $k_x $, $ m$, $Z_0$ and $\sigma_0$ are assumed to be real constants. 
Integrating  equation(\ref{Eq:Vel_MdBB_fun_x}), one obtains  the real and imaginary components of $x$ as

\begin{equation}
x_r(t)=x_{r0} + \dfrac{\hbar k_x}{m} t ,
\label{Eq:Vel_MdBB_x_r} 
\end{equation}
and 

\begin{equation}
x_i(t)=x_{i0}.
\label{Eq:Vel_MdBB_x_i}
\end{equation}

\begin{figure}[h]
\includegraphics[width=0.8\textwidth]{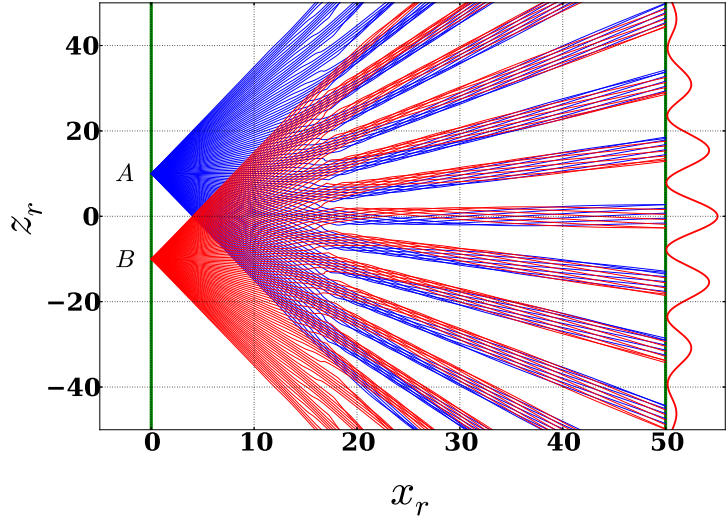}
\caption{MdBB trajectories for two interfering Gaussian wave packets, with  initial points  $x_{r0}=0 $ and $x_{i0}=0$, $z_{r0}=+10$ for slit A and $z_{r0}=-10$ for slit B. Trajectories were plotted for a uniform distribution of $z_{i0}$, ranging   from $-5$ to $+5$ for both slits A and B. Adjacent panel on the right side shows standard $\psi^{\star}\psi$ probability density in this case, along the $z$-axis ( at $x=50$ and $y=0$) on the screen.}
\label{fig:MdBB_wavepacket}        
\end{figure}

Again, we put $y_0=0$ as the initial condition for the variable $y$.
Equation (\ref{Eq:Vel_MdBB_fun_z}) was solved using fourth order Runge-Kutta  method  to obtain $z_r(t)$ and $z_i(t)$ from $ t=0$ to $t=T$. First, this equation is separated into real and imaginary parts. The Runge-Kutta method specific to  simultaneous first order differential equations, with $z_r$ and $z_i$ as the variables, is used with step-size $\Delta t=0.01$. The values for $T$, $\hbar/m$, $\sigma_0$, $k_x$, etc. are given the same values as in the previous examples. The initial conditions at $t=0$ were chosen as  $x_{r0}=0 $ and $x_{i0}=0$, $z_{r0}=Z_0=10$ for slit A and $z_{r0}=-Z_0=-10$ for slit B. In both cases,  trajectories were plotted for various values of $z_{i0}$, ranging   from $-\delta$ to $+\delta$, separated by equal intervals, with the value  $\delta =5$. Figure \ref{fig:MdBB_wavepacket}  shows plots of $x_r$ versus $z_r$, which are the projections of the complex trajectories onto the real plane. 

An advantage of the present choice of initial conditions is that on the real plane, trajectories can have their precise starting points either at slit A or slit B. The trajectories can cross each other, so that on the screen it is not possible for us to identify the slit through which a particular particle has emanated. The adjacent panel in Fig. \ref{fig:MdBB_wavepacket} shows the  $\psi^{\star}\psi$-distribution. It is easily seen that the band width obtained from the trajectories  are in good agreement with the standard values.

In addition to the crossings of trajectories emanating from holes A and B, we observe  that there are crossings between trajectories starting from the same hole. Thus we see in Fig. 3 that there are not only red/blue crossings, but also red/red and blue/blue crossings.  To see explicitly what happens in the same-color crossings, we plotted the endpoints of  trajectories on the screen, as it appears in the complex $z$-plane. When   the position of the screen is changed by changing it $x_r$-coordinate, it is realised from these patterns that the trajectories lie along spiralling, helical paths.  When  the projections of these helical paths to the real $x_rz_r$-plane are taken, the red/red, blue/blue and red/blue crossings appear to occur, though in the complex $xz$-space no crossings take place. We have observed that such same-color crossings are helpful in maintaining the interference pattern intact as the screen is moved back and forth.

 \section{Interference pattern in the MdBB approach - Stationary state}

Next we assume that the stationary dispherical wave function (\ref{eq:dispherical})  permeates the region beyond the barrier containing  the two holes, which are located at  $(0,0, \pm 10)$.   Substituting for $r_1$ and $r_2$ (respectively from equations (\ref{eq:rone}) and (\ref{eq:rtwo})) in (\ref{eq:dispherical}) and using equation (\ref{Eq:Vel_MdBB}), we  calculate the MdBB trajectories in Cartesian coordinates $(x, y, z)$.   As in the previous case,   the MdBB equations of motion is given by (\ref{Eq:Vel_MdBB}), but with $(x,y,z)$ as complex variables.
The partial derivatives to be used in these equations are, respectively, given by equations (\ref{eq:parpsid1}), (\ref{eq:parpsid3}) and (\ref{eq:parpsid2}).

In contrast  to the dBB trajectories corresponding to the dispherical state discussed in Sec. {\ref{sec:dbb_stat}, the present complex trajectories can have  starting points at ($x_{r0}=0$, $y=0$, $z_{r0}=\pm Z_0=10$), which are the precise real positions of the holes. However, one cannot take the imaginary values $x_{i0}$ and $z_{i0}$ too to be zero, for  the dispherical wave function itself is infinite at such points. Hence we shall begin plotting the trajectories from points equidistant on a circle  surrounding the holes and lying in the imaginary $xz$-plane with  radius $a$, where   $x_{i0}^2+z_{i0}^2=a^2$. The real values $x_{r0}$ and $z_{r0}$ are as stated above. The resulting trajectories for $a =15$ are shown in Fig. \ref{fig:MdBB_stationary1}.

\begin{figure}[h]
\includegraphics[width=0.8\textwidth]{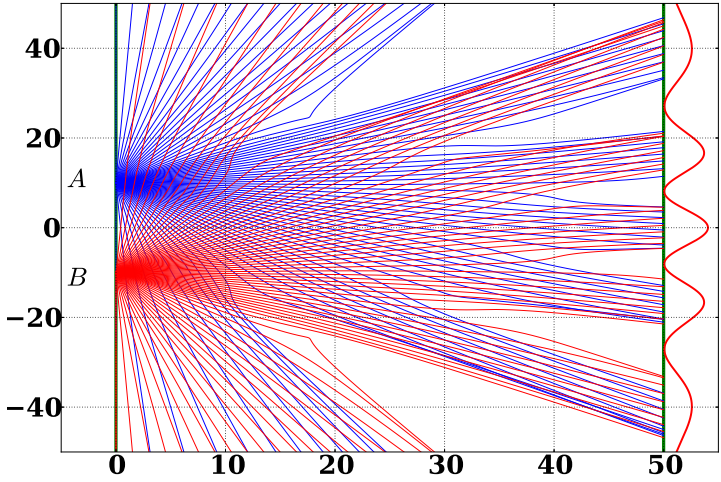}
\caption{MdBB trajectories for stationary dispherical wave function, starting from initial points with $x_{r0}=0$, $z_{r0}=\pm 10$. The imaginary values  $x_{i0}$ and $z_{i0}$ satisfy $x_{i0}^2+z_{i0}^2=a^2$ with $a =15$. Adjacent panel on the right side shows standard $\psi^{\star}\psi$ probability density in this case, along the $z$-axis (at $x=50$ and $y=0$) on the screen.}

\label{fig:MdBB_stationary1}        
\end{figure}

In the previous sections, we saw that the dBB trajectories   are  non-crossing and hence can reveal  which-way information even for the stationary states. In contrast, the trajectories shown in Fig. \ref{fig:MdBB_stationary1}, which are the projection of complex trajectories on the real $xz$-plane, can cross each other and hence cannot provide any which-way information. We   see that also the MdBB trajectories  exhibit  condensation of trajectories to high probability regions as the particles move on, even when they  cross each other and has no which-way information. Interestingly, the same color crossings, as observed for wave packets in the previous case of MdBB trajectories, can be observed here also. We see that trajectories from either holes come together and move along helical paths in the complex space. Again, this helps to maintain the interference pattern as the screen is moved back and forth.

\section{Probability distribution on the screen}  \label{sec:Prob_Scr}
 
In all the above cases of plotting the dBB and MdBB trajectories,  we find very good agreement between trajectory formalism and the standard one, with regard to the band width of the interference pattern. We observe that very few trajectories (either red or blue) reach the screen, where the probability is expected to be small. Similarly, both the red and blue trajectories accumulate at  regions where the probability  $\psi^{\star}\psi$ is high.  For the    MdBB trajectories in the stationary state,  an attempt was also made to evaluate the probability density on the screen, on the basis of the number density of trajectories reaching there. Clearly the number of trajectories reaching a certain region on the screen depends on the initial distributions of starting points. We anticipate that if this initial distribution is chosen according to a $\psi^{\star}\psi$-distribution, the final distribution on the screen, obtained by counting the number of trajectories reaching each small segments on it,  agrees  with the standard distribution. As in the case of plotting these trajectories in the above section, the starting points are chosen as the real positions of the holes and lying on a circle in the imaginary $xz$-plane with  radius $x_{i0}^2+z_{i0}^2=a^2$. But instead of equidistant points, we now choose the initial distribution of points along this curve in such a way as to obey $\psi^{\star}\psi$. Even though this is not  an exhaustive distribution of initial points near the holes, we could get very good agreement with the standard intensity distribution on the screen, as can be seen from Fig.\ref{fig:distribution}. Here, the  normalised distribution obtained by counting the trajectories is plotted, along with the standard distribution in this case. Thus we see that one can obtain not only the band width, but also the probability distribution and hence the intensity distribution  on the screen, using the MdBB trajectory representation.

\begin{figure}[h]
\includegraphics[width=0.8\textwidth]{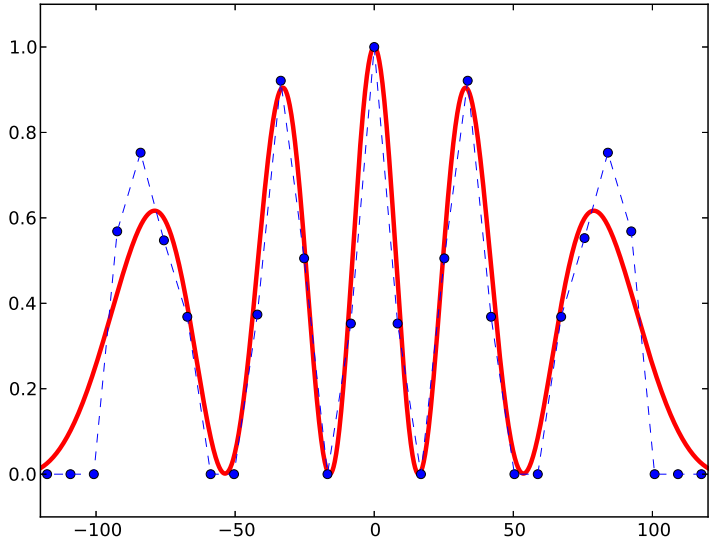}
\caption{Comparison of standard $\psi^{\star}\psi$ distribution (red thick line) and  the distribution obtained by counting the MdBB trajectories (blue dotted line) on the screen, for the dispherical wave function discussed in Sec. \ref{sec:Prob_Scr}}
\label{fig:distribution}        
\end{figure}

\section{Summary}
 In the quantum folklore,  interference  in a double slit experiment is  the  phenomenon which involves the most important, or perhaps the only, mystery in quantum mechanics \cite{feynman}. In this paper, we have presented an analysis of this experiment, based on the dBB and MdBB quantum trajectory representations. Our attempt was to find whether  both  the MdBB and dBB formalisms can provide   satisfactory explanation of the phenomenon, even for stationary states. Also, we aimed to investigate these trajectories regarding knowledge of which-way information, that may help one to identify the slit through which a particle emanates. 
 
In the original viewpoint of  de Broglie's  wave-particle duality,  physical systems have  wave and particle nature, so that they can be   described    using a `both particle and wave' representation such as the `pilot wave'. de Broglie strived to show that if a phenomenon can be explained as due to wave motion, it is possible to explain it as due to particle motion as well, even if that requires some modification to Newtonian particle  mechanics. In the Young's double slit experiment, interference  is traditionally demonstrated  as resulting from the superposition of two stationary coherent  waves emanating from the slits. In fact, this is the classic example used by   generations of physicists to  understand the phenomenon. In the literature,  the dBB trajectory formalism has   demonstrated interference as arising from particle motion, but so far only for   spreading  wave packets. In this paper, first we have  demonstrated that interference of two stationary spherical waves  can also  result from particle trajectories, in the dBB formalism. This brings this scheme closer to  de Broglie's principle of wave-particle duality. However, it was noted that  all the dBB trajectory explanations in the double-slit experiment allow which-way information regarding particle motion. 
  
 We have also analysed the problem in  the alternative MdBB approach, that gives an entirely different set of trajectories.  Our analysis has shown that the MdBB trajectories are capable of providing exactly  the same  interference pattern on the screen as that obtained in standard quantum mechanics and the dBB approach, for both the wave packet case and the stationary state case. We also obtained the result  that these MdBB trajectories  cross each other and hence can explain quantum interference while not givng away any which-way information. This happens because  the MdBB velocity field is more general than the dBB velocity. To be specific, the MdBB velocity field (\ref{Eq:Vel_MdBB}) is defined over the entire complex plane and has real and imaginary components, but the dBB velocity  (\ref{eq:vel_dBB}) is only the real part of it.  In this work, we have also made a trajectory-based calculation of the probability density on the screen for the MdBB scheme and found that when the  distribution of the starting points of trajectories are based on the $\psi^{\star}\psi$-probability density,  the  distribution of their end-points on the screen also obeys it. This is accomplished even for the case of stationary states with no probability flow.
\begin{acknowledgements}
We wish to thank an anonymous Reviewer for helping to point out the important feature of same-color crossings in the MdBB trajectory plots. We also thank Professor K. Babu Joseph for useful discussions. 
\end{acknowledgements}



\end{document}